\documentclass{article}
\usepackage[utf8]{inputenc}
\usepackage[margin=0.75in]{geometry}
\usepackage{bm}
\usepackage{graphics,graphicx,pgf,natbib,amsmath,amsthm,amssymb,url}

\numberwithin{equation}{section}
\newtheorem{Defn}{Definition}
\newtheorem{thm}{Theorem}
\newtheorem{result}{Result}
\newtheorem{remark}{Remark}

\newcommand{\bfSigma}{\mathbf{\Sigma}}

\title{\textbf{Sparse Portfolio Selection via Bayesian Multiple Testing}}
\author{}
\date{Sourish Das \footnote{ The Research of Sourish Das is supported by Bill \& Melinda Gates Foundation grant, TATA Trust grant and Infosys Foundation grant to CMI. {\it{Corresponding author}} is Rituparna Sen Email: rsen@isibang.ac.in \textbf{Accepted in Sankhya-B.}} \\
Chennai Mathematical Institute, Chennai, INDIA. \\
\vspace{0.5cm}  and \\\vspace{0.5cm} 
Rituparna Sen \\ Indian Statistical Institute, Applied Statistics Unit, Bengaluru, INDIA. \\\vspace{0.75cm}
 21 July 2020}

\begin{document}

\maketitle
\begin{abstract}
We present Bayesian portfolio selection strategy, via the $k$ factor asset pricing model. If the market is information efficient, the proposed strategy will mimic the market; otherwise, the strategy will outperform the market. The strategy depends on the selection of a portfolio via Bayesian multiple testing methodologies. We present the ``discrete-mixture prior" model and the ``hierarchical Bayes model with horseshoe prior."  We define the oracle set and prove that asymptotically the Bayes rule attains the risk of Bayes oracle up to $O(1)$. Our proposed Bayes oracle test guarantees statistical power by providing the upper bound of the type-II error.  Simulation study indicates that the proposed Bayes oracle test is suitable for the efficient market with few stocks inefficiently priced. The statistical power of the Bayes oracle portfolio is uniformly better for the $k$-factor model ($k>1$) than the one factor CAPM. We present an empirical study, where we consider the 500 constituent stocks of S\&P 500 from the New York Stock Exchange (NYSE), and S\&P 500 index as the benchmark for thirteen years from the year 2006 to 2018. We show the out-sample risk and return performance of the four different portfolio selection strategies and compare with the S\&P 500 index as the benchmark market index. Empirical results indicate that it is possible to propose a strategy which can outperform the market. All the \texttt{R} code and data are available in the following GitHub repository \url{https://github.com/sourish-cmi/sparse_portfolio_Bayes_multiple_test}.
\end{abstract}


\section{Introduction}

Markowitz portfolio theory \citep{Markowitz.1952} in finance analytically formalizes the risk-return tradeoff in selecting optimal portfolios. An investor allocates the wealth among securities in such a way that the portfolio guarantees a certain level of expected returns and minimizes the `risk' associated with it. The variance of the portfolio return is quantified as the risk.

Markowitz portfolio optimization is very sensitive to errors in the estimates of the expected return vector and the covariance matrix, see, e.g., \citep{Fan.Zhang.Yu.2012}. The problem is severe when the portfolio size is large. Several techniques have been suggested to reduce the sensitivity of the Markowitz optimal portfolios. One approach is to use a James-Stein estimator for means (i.e., expected return) \citep{Chopra.Ziemba.1993} and shrink the sample covariance matrix \citep{Ledoit.Wolf.2003, Das.Halder.Dey.2016}. Still, the curse of dimensionality kicks-in for a typically large portfolio (like mutual fund portfolio) and the procedure underestimates the risk profile of the portfolio \citep{EK.2010}. Hence dimension reduction is an essential requirement for robust portfolio selection.

Another reason to avoid large portfolios is to reduce transaction costs. As the number of assets in the portfolio increase, the transaction cost increases. For example a common model for transaction costs is linear plus fixed as presented in \citep{lobo2007}. This means there is a fixed cost in investing in an asset, plus an amount proportional to the investment amount. Thus, if the total value of the portfolio remains same, the linear part of the transaction cost will be same. However the fixed part will be proportional to the number of assets in the portfolio.

In this paper, in order to address the dimension reduction problem, we take the alternative route for portfolio selection, which goes through the capital asset pricing model (CAPM), see, e.g. \citep{Sharpe.1964, Linter.1965, Black.1972}. Although CAPM fails to explain several features, including rationality of the investors; it has become a standard tool in corporate finance \citep{Goyal.2012}. The CAPM splits a portfolio return into \emph{systematic} return and \emph{idiosyncratic} return and models it as the linear regression of portfolio's `\emph{risk premium}' (aka. `\emph{excess returns}') on the market's `\emph{risk premium}'. If the value of the intercept in this regression is zero, then the asset is fairly valued.  If the intercept is zero and the slope is one, then the asset behaves very similarly to the market, and there is no gain in including it in the portfolio in addition to the market. The Fama and French Three-Factor Model, see \citep{Fama.1993}, is an asset pricing model that expands on the CAPM by adding size risk and value risk factors to the market risk factor in CAPM. In general, for a $k$ factor model, the asset will behave very similar to the market if the coefficient of the market is one and all other coefficients are zero. For most assets under consideration, this is the case. Thus the objective is to select such assets that have different behavior than the market and construct the portfolio based on those and the market. We present the problem of selecting such assets as multiple testing problems, under sparsity when the number of assets is high.

Bayesian methods are proposed to test the restriction imposed in CAPM that the intercepts in the regression of `\emph{risk premium}' on the `\emph{market risk premium}' are equal to zero, see \citep{Shanken.1987, Harvey.Zhou.1990}. Shanken's methodology relies on the prior induced on functions of intercept and sampling distribution of $F$-statistics, see \citep{Shanken.1987}. Harvey and Zhou \citep{Harvey.Zhou.1990} proposed a full Bayesian specification of the CAPM test with diffuse prior and conjugate prior structure. Black and Litterman  \citep{Black.Litterman.1992} presented an informal Bayesian approach to economic views and equilibrium relations. The existing methodology concentrates on the test for intercept (or $\alpha$) only. In our proposed method we present a joint test for both intercept and slope, (i.e., $\alpha$ and $\beta$) for the CAPM and a multivariate test for the general $k$ factor model. We present the joint test in the next section.

Rest of the paper is organized as follows. In Section \ref{Section_Proposed_Strategy},  we propose a portfolio selection strategy based on sparsity. In Section \ref{Section_Methodology} we present the discrete-mixture prior model and hierarchical Bayes model with half-Cauchy distribution on the scale parameters. In Section \ref{Section_Asymptotic_Optimality} we derive the results of the asymptotic Bayes optimality for the multiple testing methodology proposed in section \ref{Section_Methodology}. In Section \ref{Section_Simulation}, we conduct a simulation study. In Section \ref{Section_Empirical}, we present the empirical study based on data from the New York Stock Exchange (NYSE). We conclude the paper with a discussion in Section \ref{Section_Discussion}.

\section{Strategy}\label{Section_Proposed_Strategy}
In this section, we present the strategy for portfolio selection. We discuss that when the maximum weight of an asset in  a large portfolio is small, the idiosyncratic risk of the portfolio washes out (see the Result (\ref{result_idiosyn_wash_out}) in Appendix). This feature holds for both the CAPM, and its extension to the $k$-factor model. In this set-up the portfolio return can be mostly explained by market movements and corresponding regression coefficients, popularly known as the `\textbf{portfolio's beta}'.

In matrix notation, the CAPM is presented as follows:
\begin{eqnarray}\label{eqn_CAPM}
\bm{r}=\bm{XB}+\bm{\epsilon},
\end{eqnarray}
where
$\bm{r}=\left((r_{i,j})\right)_{n\times P}$
is the matrix of excess return over the risk free rate for $P$ many assets that are available in the market over $n$ days; $\bm{XB}$ is the systematic return due to market index, where
\begin{equation}\label{eqn:CAPM}
\bm{X}=\left((\bm{1} ~~\bm{r}_m)\right)_{n\times 2}
\end{equation}
is the design matrix with the first column being the unit vector or the place holder for intercept and the second column being the vector of excess returns of the market index over the risk free rate;
$$
\bm{B}=\left(\begin{array}{cccc}
 \alpha_1 & \alpha_2 & \hdots & \alpha_P\\
 \beta_1  & \beta_2 & \hdots & \beta_P\\
\end{array}\right)_{2\times P}.
$$
If the market is efficient then according to \citep{Sharpe.1964,Linter.1965,Black.1972}, the intercepts $\alpha_i=0~~\forall i=1\cdots P$ and $\beta_i$ is the measure of systematic risk due to market movement;
$\bm{\epsilon}=\left((\epsilon_{i,j})\right)_{n\times P}$
is the idiosyncratic return of the asset.
\begin{remark}
In (\ref{eqn_CAPM}), if a portfolio is constructed based on $\tilde{P}$  many assets, all of which have $\alpha=0$ and $\beta=1$, then portfolio return will mimic the market return.
\end{remark}

The Fama and French Three-Factor Model has similar representation as CAPM given by equation (\ref{eqn:CAPM}) with
$$
\bm{X}=\left((\bm{1} ~~\bm{r}_m ~~SMB ~~HML)\right)_{n\times 4}
$$
where $SMB$ stands for ``Small Minus Big'' in terms of market capitalization and $HML$ for ``High Minus Low'' in terms of book-to-market ratio. They measure the historic excess returns of small caps over big caps and of value stocks over growth stocks. These factors are calculated with combinations of portfolios composed by ranked stocks and available historical market data. Also, now
$$
\bm{B}=\left(\begin{array}{cccc}
 \alpha_1 & \alpha_2 & \hdots & \alpha_P\\
 \beta_1  & \beta_2 & \hdots & \beta_P\\
 b^s_1 & b^s_2 &\hdots &b^s_P\\
 b^v_1 & b^v_2 &\hdots &b^v_P\\
\end{array}\right)_{4\times P}.
$$

In general one can consider a $k$-factor model with $\bm{X}$ being $n\times (k+1)$ dimensional, where the first column of $\bm{X}$ is constant, the second column is the market return and the remaining are other $(k-1)$ suitable factors. $\bm{B}$ is $(k+1)\times P$ dimensional and we shall denote it as \begin{equation}\bm{B}=(\theta_1 \cdots \theta_P)\quad \mathrm{where} \quad\theta_i=(\alpha_i\quad \beta_i \cdots b_i^{(1)}\quad\cdots b_i^{(k-1)})^T.\label{eqn:theta}\end{equation}

\begin{remark}
In (\ref{eqn_CAPM}), for the $k$ factor model, if a portfolio is constructed based on $\tilde{P}$  many assets all of which have  $\alpha=0$, $b^{(j)}=0 ~\forall~ j\in\{1, \cdots, k-1\}$ and $\beta=1$, then the portfolio return will mimic the market return.
\end{remark}

The covariance of $r_i$ is $\bfSigma$, which can be decomposed into
\begin{eqnarray}\label{eqn_cov_matrix}
\bfSigma=\bm{B}^T\bfSigma_X\bm{B}+\bfSigma_{\epsilon},
\end{eqnarray}
where $\bfSigma_{\epsilon}=\mathrm{diag}(\sigma_1^2,\sigma_2^2,\hdots,\sigma_P^2)$ and $\bfSigma_X$ is the covariance matrix of $\bm{X}$. Let us consider a portfolio $\mathbf{w}=\{w_1,\hdots,w_P\}$, where  $0\leq w_i , i=1,\hdots,P, \sum_{i=1}^Pw_i=1$.
We show in the Appendix that, if $$P\rightarrow \infty,\quad M_{\omega:P}:=\max\{\mathbf{w}\} \rightarrow 0\quad \mathrm{and}\quad \sigma_{max}^2=\max\{\bfSigma_{\epsilon}\}<\infty$$, 
$$\mathrm{then} \quad
\lim_{P\rightarrow \infty} \mathbf{w}' \bfSigma_{\epsilon} \mathbf{w} =0.
$$

\begin{remark}
Thus we can select $\tilde{P}(\ll P)$ many assets for the portfolio (out of $P$ many assets available in the market), such that the idiosyncratic risk is washed out, i.e., for all $\delta >0,~ \exists ~\tilde{P}_{\delta}$, such that for $\tilde{P}>\tilde{P}_{\delta}$,
$$
 ||\mathbf{w}_{\tilde{P}}' \bfSigma_{\tilde{P}} \mathbf{w}_{\tilde{P}}|| <\delta;
$$
and portfolio return is mostly explained by $\theta$ only. Note that here $\tilde{P}$ is the effective size of the portfolio.
\end{remark}

\begin{remark}
\textbf{Oracle Set}: Suppose the market is not efficient and there are $q$ many assets whose $\alpha > 0$, where $q < \tilde{P}\ll P$. Let us call this set $A_q$. We can construct a portfolio with $\tilde{P}$ many assets, such that
\begin{equation}
\mathrm{P}(A_q \subset B_{\tilde{P}})\geq 1-\eta,\label{eqn_oracle_inclusion_prob}
\end{equation}
where $B_{\tilde{P}}$ is the set of assets in the portfolio, $0<\eta<1$ and $\mathbf{w}'_{\tilde{P}} \bfSigma_{\tilde{P}} \mathbf{w}_{\tilde{P}}
< \delta$. Note that $\mathbf{w}'_{\tilde{P}} =\{\omega_i : \max_{i=1(1)\tilde{P}}{||\omega_i||}<\delta\}$ and $\bfSigma_{\tilde{P}}=diag(\sigma_1^2,\hdots,\sigma_{\tilde{P}}^2)$ is
the covariance matrix of the idiosyncratic returns.

\noindent \textbf{Example}: Suppose the market
consists of $P=2000$ stocks and a portfolio manager wants to build the
portfolio with $\tilde{P}=100$ stocks. If $q=5$ many stocks
are available with $\alpha_j >0, j=1,2...,5$,
then the portfolio manager would like to build a portfolio,
such that $A_{q=5}$ is the subset of manager's selected
portfolio $B_{\tilde{P}=100}$. In other words, the manager wants to
build her portfolio in such a way that she does not want to miss out
the set of five under-valued stocks $A_{q=5}$. That is, she wants to
employ a statistical methodology, where $\mathrm{P}(A_{q=5} \subset
B_{\tilde{P}=100})$ would be very high. Note that if the market is
efficient, then $A_q$ will be a null set.
\end{remark}
The problem reduces to identifying the oracle set $A_q$. Essentially, it is a multiple testing problem, where we select those stocks in the portfolio $B_{\tilde{P}}$ for which we reject the following null hypothesis:
\begin{equation}\label{eqn:hypothesis}
H_{0i}: \left(\begin{array}{cc}\alpha_i\\\beta_i\\b^{(1)}_i\\\vdots\\b^{(k-1)}_i\end{array}\right) =
\left(\begin{array}{cc}0\\1\\0\\\vdots\\0\end{array}\right)~~~ vs. ~~ H_{Ai}:
\left(\begin{array}{cc}\alpha_i\\\beta_i\\b^{(1)}_i\\\vdots\\b^{(k-1)}_i\end{array}\right) \ne
\left(\begin{array}{cc}0\\1\\0\\\vdots\\0\end{array}\right),~~~ i=1\hdots P.
\end{equation}
We would like to define an optimal test rule, such that (\ref{eqn_oracle_inclusion_prob}) is satisfied.

Here, the structure of the multiple testing problem is very different, compared to typical multiple testing problems in the literature \cite{Carvalho.Polson.Scott.2009,Carvalho.Polson.Scott.2010,Datta.Ghosh.2013}, which are mainly motivated from genome wide association study. In the next section, we present the Bayesian methodology to identify $B_{\tilde{P}}$.

\section{Methodology}\label{Section_Methodology}
In this section, we propose Bayesian methodologies for testing the null hypothesis $\theta_i=\mu_0 :=(0, 1,0,\cdots, 0) ~\forall i=1\cdots P$, where $\theta_i$ is defined in equation (\ref{eqn:theta}). First we propose the discrete mixture prior and then we propose the hierarchical Bayes model.

For each $i=1\hdots P$, under the assumption of multivariate normality of $(\epsilon_{ij}, j=1\cdots n)$,  the least squares estimator
$\hat{\theta}_i=(\hat{\alpha}_i,\hat{\beta}_i ,\hat{b}^{(1)}_i,\cdots, \hat{b}^{(k-1)}_i)^T$, is the MLE of $\theta_i$. This is also sufficient statistics and the sampling distribution is $\hat{\theta}_i \sim N_{k+1}\bigg(\theta_i~,\sigma_i^2\Sigma_X^{-1}\bigg)$
where $\Sigma_X=X^TX$. For the rest of this section, the underlying model is the $k$ factor model.

\subsection{Bayes oracle with discrete-mixture prior}\label{Section_DMP}
We propose to use a discrete mixture prior, commonly known as the `spike and slab' prior introduced by \citep{mitchell1988bayesian}. The prior puts probability $1-p$ on $\theta_i=\mu_0$ and $p$ on an absolutely continuous alternative, denoted by $M_a$, as $\left[\theta_i|\Lambda_{0} \right]\sim N_{k+1}(\mu_0, \Lambda_{0}^{-1})$. Unconditionally, $\theta_i$'s are independently distributed as
\begin{eqnarray*}
\theta_i\sim (1-p)\delta_{\mu_0}+p N_{k+1}(\mu_0,\Lambda_{0}^{-1}),
\end{eqnarray*}
where $\delta_{\mu_0}$ is the degenerate distribution at $\mu_0$. For each $i$, the null hypothesis implies prior mean of $\mathrm{E}(\alpha_i)=0$, $\mathrm{E}(\beta_i)=1,\mathrm{E}(b^{(1)}_i)=0,\cdots, \mathrm{E}(b^{(k-1)}_i)=0$. The parameter $p$ is often known as the sparsity parameter. As $p\rightarrow 0$ the model becomes a sparse model and as $p\rightarrow 1$ the model is known as dense model. This implies that the marginal distribution of $\hat{\theta}_i$ is the scale mixture of normals, that is
\begin{equation}\label{eqn_marginal}
\hat{\theta}_i\sim (1-p)N_{k+1}(\mu_0, \sigma^2_i\Sigma_X^{-1})+pN_{k+1}(\mu_0, \sigma^2_i\Sigma_X^{-1}+\Lambda_{0}^{-1}).
\end{equation}
 The conditional posterior distribution under the alternative is
\begin{equation}\left[\theta_i\mid\sigma_i,\Lambda_{0},\hat{\theta_i}\right]\sim N_{k+1}(\mu_{ni}, \Lambda_{ni}^{-1}),\nonumber\end{equation}
\begin{eqnarray*}\mathrm{where}\quad
\Lambda_{ni}&=&\Lambda_{0}+\frac{\Sigma_X}{\sigma^2_i}
~~\mathrm{ and }\quad\mu_{ni}=\Lambda_{ni}^{-1}\left(\Lambda_{0}\mu_{0}+\frac{\Sigma_X}{\sigma_i^2}\hat{\theta_i}\right).\end{eqnarray*}
Under a sparse mixture model, the Bayes oracle has the rejection region $\mathcal{C}$ on which the Bayes factor exceeds $\frac{(1-p)\delta_0}{p\delta_A}$, where $\delta_0$ and $\delta_A$ are the losses associated with type I and type II errors, see \citep{Bogdan.2011}. In this case, the Bayes factor can be computed as  \begin{eqnarray}&\left(\mathrm{det}(I-Q_i)\right)^{1/2}\exp(\frac{S_i}{2}),&\nonumber\\&\textrm{where}\quad S_i = (\mu_{ni}-\mu_0)^T\Lambda_{ni}(\mu_{ni}-\mu_0)&\nonumber\\& ~~\textrm{ and }\quad Q_i = \frac{X}{\sigma_i}\Lambda_{ni}^{-1}\frac{X}{\sigma_i}^T&\label{eqn:Q}\end{eqnarray}
The optimal rule is to reject $H_{0i}$ if \begin{equation}\label{eqn_oracle}S_i\ge c_i^2=-\log\left(\mathrm{det}(I-Q_i)\right)+2\log(f\delta),\end{equation} \begin{equation}\label{eqn:fd}\mathrm{where}\quad f=\frac{(1-p)}{p}\quad\mathrm{and}\quad \delta=\frac{\delta_0}{\delta_A}.\end{equation} We call this rule {\it Bayes oracle} since it makes use of unknown parameters $p,\Lambda_{0}$ and cannot be attained in finite samples. The posterior inclusion probability is \begin{equation*}\tilde{\pi_i}=Pr(M_a|D)=\left(1+\frac{1-p}{p}\exp(-\frac{S_i}{2})\right)^{-1}, \quad \textrm{by Bayes' theorem.}\end{equation*}
Under symmetric loss, the rejection region coincides with $\tilde{\pi_i}>1/2.$
The test statistic involves
\begin{equation*}
S_i=\frac{(r_i-X\mu_0)}{\sigma_i}^TQ_i \frac{(r_i-X\mu_0)}{\sigma_i}.
\end{equation*}
Under the null hypothesis,  $(r_i-X\mu_0)/\sigma_i\sim\mathcal{N}(0,I)$. Thus $S_i$ is a quadratic form in multivariate normal with $Q_i$ symmetric non-idempotent of rank $k+1$. The distribution is weighted sum of central $\chi^2$ random variables of 1 degree of freedom (df) with weights being the eigenvalues of matrix $Q$, see \citep{Vuong.1989}. In summary,
\begin{equation*}
S_i\sim \sum_{j=1}^{k+1}\lambda_{ji}\chi^2_j,
\end{equation*}
where $\lambda_{ji}$ are the $k+1$ non-zero eigenvalues of $Q_i$ and $\chi^2_j$ are independent central $\chi^2$ random variables with 1 df. The distribution denoted by $M_{k+1}(.;\lambda)$ is well studied, see for e.g. \citep{Solomon.1977}.
The probability of type I error  is
\begin{equation*}
t_{1i}=P(\sum_{j=1}^{k+1}\lambda_{ji}\chi^2_j\ge c_i^2).
\end{equation*}
Under the alternative hypothesis the marginal distribution of the data is,  $(r_i-X\mu_0)\sim\mathcal{N}(0,\sigma_i^2A_i)$ where $A_i=(I-Q_i)^{-1}$. The distribution of $S_i$ is weighted sum of central $\chi^2$ random variables of 1 df with weights being the eigenvalues of matrix $A_i^{1/2}Q_iA_i^{1/2}=(I-Q_i)^{-1}-I$. Eigenvalues of  this matrix are $\frac{\lambda_{ji}}{1-\lambda_{ji}}$, where $\lambda_{ji}, j=1, \cdots,k+1$ are the non-zero eigenvalues of $Q_i$ as before.
The probability of type II error is
\begin{equation*}
t_{2i}=P\left(\sum_{j=1}^{k+1}\frac{\lambda_{ji}}{1-\lambda_{ji}}\chi^2_j\le c_i^2\right).
\end{equation*}
Under additive loss function, the Bayes risk of the Bayes oracle is
\begin{equation*}
R_{\mathrm{opt}}=(1-p)\delta_0\sum_{i=1}^Pt_{1i}+p\delta_A\sum_{i=1}^Pt_{2i}.
\end{equation*}

\subsection{Hierarchical Bayesian approach}
We present the hierarchical Bayesian approach for the  $k$-factor model, in the spirit of \cite{Gelfand.1990}, which is as follows.
\begin{eqnarray*}
r_{ti} &\sim& N(X_t\theta_i,\sigma_i^2),\quad t=1,\cdots,n, \quad \mathrm{and}\quad i=1,\cdots,P.\\
\theta_i|\Lambda&\sim& N_{k+1}(\theta_0,\tau^2\Lambda^{-1}),
\end{eqnarray*}
where $\theta_i=(\alpha_i,\beta_i, b^{(1)}_i,\cdots,b^{(k-1)}_i)$, and $\theta_0=(\alpha_0,\beta_0, b^{(1)}_0,\cdots,b^{(k-1)}_0)$
\begin{eqnarray*}
\sigma_i^2 &\propto& InvGamma\bigg(\frac{\nu_0}{2},\frac{\nu_0}{2}\bigg),\\
\Lambda&\sim& Wishart\Big( (\rho R)^{-1},\rho\Big),\\
\theta_0 &\sim& N_{k+1}(\mu_0,C),\\
\tau^2&\sim& C^+(0,1)
\end{eqnarray*}
where $R$ is the prior scale matrix, $\rho$ is prior degrees of freedom of the Wishart distribution, $\mu_0=(0,1,0,\cdots,0)^T$. We assumed $C$ to be the identity matrix of order $k+1$. Here $C^+(0,1)$ denotes the half-Cauchy distribution with location parameter 0 and scale parameter 1 with corresponding pdf as
$$f(\tau)=\frac{2\mathrm{I}(\tau>0)}{\pi(1+\tau^2)}.$$
The scale parameter $\tau$ plays a crucial role in controlling the shrinkage behavior of the estimator. It is known as ``global shrinkage parameter" \cite{Datta.Ghosh.2013}, as it adjusts to the overall sparsity in the data. $\Lambda$ behaves as local shrinkage parameter. Note that we consider the half-Cauchy prior \citep{Gelman2006} over $\tau$. \cite{Carvalho.Polson.Scott.2009,Carvalho.Polson.Scott.2010} showed that such prior specification is suited for high-dimension sparse solution problem and named it as `horseshoe prior'.  The posterior probability of $\tau$ is concentrated near zero when the data is sparse ($p \rightarrow 0$). We define $$\bar{\theta}=P^{-1}\sum_{i=1}^P\theta_i,$$
$$V=(P\Lambda+C^{-1})^{-1}.$$
The Gibbs sampler for $\theta_i$, $\sigma_i^2$, $\Lambda$ and $\theta_0$ is straight forward as
$$[\theta_i|X,r_i, \mu_c, \Lambda,\sigma_i^2 ]\sim N_{k+1}(m_i,\Sigma_i^{-1}),$$
where $\Sigma_i=\frac{X^TX}{\sigma_i^2}+\frac{\Lambda}{\tau^2}$ and $m_i=\Sigma_i^{-1}(\frac{X^Tr_{i}}{\sigma_i^2}+\frac{\Lambda}{\tau^2} \mu_0).$
$$[ \sigma_i^2|n_i,r_i,\theta_i ]\sim InvGamma\bigg(\frac{\nu_0 + n_i}{2},\frac{\nu_0 + (r_{i}-X\theta_i)^T(r_{i}-X\theta_i)}{2}\bigg),$$
$$[ \Lambda | \theta_i,P, R, \rho,\mu_c] \sim Wishart\Big\{ \big(\sum_{i=1}^P(\theta_i-\mu_c)(\theta_i-\mu_c)^T+\rho R\big)^{-1},P+\rho\Big\},$$
$$[\theta_0|V,P,\Lambda,C,\mu_0] \sim N_{k+1}\Big(V(P\Lambda\bar{\theta}+C^{-1}\mu_0),V\Big).$$
We implemented a Metropolis-Hastings update for $\tau$. 

\section{Asymptotic Optimality}\label{Section_Asymptotic_Optimality}
The notion of `Asymptotic Bayes Optimality under Sparsity' (ABOS) was introduced in \citep{Bogdan.2011}. This has been extended to show optimality of one-group models in \citep{Datta.Ghosh.2013}. In particular, it is shown that if the global shrinkage parameter $\tau$ of the horseshoe prior is chosen to be the same order as $p$, then the natural decision rule induced by the horseshoe prior attains the risk of the Bayes oracle up to O(1) with a constant close to the constant in the oracle. There have been several studies following this in the one-parameter setting. Here we extend these results to the multi-parameter setting. The asymptotic framework that we work under is motivated by \citep{Bogdan.2011}. For extending the result to the ($k+1$)-dimensional case, we need the following assumption.

\noindent \textbf{Assumption} (A): A  sequence  of  parameter vectors $\{\gamma_t=(p_t,\Lambda_{0t},\sigma_{it},\delta_t);t\in\{1,2,\cdots\}\}$
satisfies  this  assumption  if  it fulfills the following conditions: \begin{eqnarray*}p_t&\rightarrow& 0,\\ \sigma_{it}^2\Lambda_{0t}&\rightarrow &0,\\ v_t:=(\mathrm{det}(I-Q_{it}))^{-1}f^2_t\delta^2_t&\rightarrow&\infty,\\ u_t\log(v_t)&\rightarrow& C\in(0,\infty),\end{eqnarray*} where $Q_i$ is defined in equation (\ref{eqn:Q}) and $f$ and $\delta$ in equation (\ref{eqn:fd}).\\ $\lambda_{ji},~j=1,\cdots,k+1$ are the non-zero eigenvalues of $Q_i$\\ and $u_t :=\left({\prod_{j=1}^{k+1}(1-\lambda_{jit})}\right)^{1/(k+1)}$.

\begin{thm}\label{Thm_T1_error_goes_to_0}
Under assumption A, $t_{1i}\rightarrow 0$ and $t_{2i}\rightarrow P(\chi^2_{(k+1)}\le C)$. In particular, for $k=1, t_{2i}\rightarrow 1-e^{-C/2}$.
\end{thm}
\begin{proof}It has been seen in section \ref{Section_DMP} that, under the null hypothesis, $S_i$ is a linear combination of $k+1$ independent $\chi^2_1$ random variables with weights $\lambda_{ji},~j=1,\cdots,k+1$ the non-zero eigenvalues of $Q_i$.\\
Under assumption A, $v_t\rightarrow\infty$. So, for the last part of the assumption to hold, each $\lambda_{ji},~j=1,\cdots,k+1$  must converge to 1 as $t\rightarrow\infty$. Hence $S_i\Rightarrow\chi^2_{k+1}$, where $\Rightarrow$ denotes convergence in distribution.\\
By the assumption of $v_t\rightarrow\infty$, we have  $c_i^2\rightarrow\infty$ and hence $t_{1i}\rightarrow 0$.

Under the alternative hypothesis  $S_i$ is a linear combination of $k+1$ independent $\chi^2_1$ random variables with weights $\lambda_{ji}/(1-\lambda_{ji}),~j=1,\cdots,k+1$. Under assumption A,  $\lambda_{ji},~j=1,\cdots,k+1$ converge to 1. Hence $u_t S_i\Rightarrow\chi^2_{k+1}$. Also, $u_t\log(v)\rightarrow C$. Hence $t_{2i}\rightarrow P(X\le C)$ where $X$ is a $\chi^2_{k+1}$ random variable, hence the result.
\end{proof}
\begin{remark}\label{Thm_Bayes_Oracle_Risk} From the above theorem, we can conclude that under the Bayes oracle, the risk takes the form $R_{\mathrm{opt}}=Pp\delta_AP(\chi^2_{(k+1)}\le C).$\end{remark}
\begin{Defn} Consider a sequence of parameters $\gamma_t$ satisfying Assumption A. We call a multiple testing rule ABOS for $\gamma_t$ if its risk $R$ satisfies \begin{equation*}\frac{R}{R_{\mathrm{opt}}}\rightarrow 1\quad \mathrm{as} \quad t\rightarrow\infty\end{equation*}\end{Defn}
We propose an alternative test with rejection region $\tilde{S}_i>c_i^2$, where $c_i^2$ is as defined in equation (\ref{eqn_oracle}) and \begin{equation*}\tilde{S}_i=\frac{(r_i-X\mu_0)}{\sigma_i}^TX(X^TX)^{-1}X^T \frac{(r_i-X\mu_0)}{\sigma_i}.\end{equation*}
The advantage of $\tilde{S}_i$ is that it does not depend on $\Lambda_0$. We show that this new test is ABOS.
\begin{thm}\label{Thm_Si_Tilde_ABOS}
The test that rejects $H_0$ when $\tilde{S}_i>c_{it}^2$ is ABOS if and only if $c_{it}\rightarrow\infty$ and $c_{it}^2u_t\rightarrow C$.
\end{thm}
\begin{proof}
Under $H_0$, $$\tilde{S}_i=\frac{(r_i-X\mu_0)}{\sigma_i}^T\tilde{Q} \frac{(r_i-X\mu_0)}{\sigma_i}$$ is a quadratic form in standard multivariate normal with $\tilde{Q}=X(X^TX)^{-1}X^T$ symmetric idempotent of rank $k+1$. 
Hence $\tilde{S}_i\sim\chi^2_{k+1}$ and $t_{1i}=P(\tilde{S}_i>c_{it}^2)\rightarrow 0$ if and only if $c_{it}\rightarrow\infty$.
Under the alternative, $$Z_i=\frac{(r_i-X\mu_0)}{\sigma_i}^TA_i^{-1/2}$$ is standard normal and $\tilde{S}_i=Z_i^TA_i^{1/2}\tilde{Q} A_i^{1/2}Z.$ So $(\mathrm{det}(A_i))^{-1/(k+1)}S_i\Rightarrow\chi^2_{k+1}$.  Also, $\mathrm{det}(A_i)=\prod_{j=1}^{k+1}(1-\lambda_{ji})$. 
The ABOS property is now established using the remark \ref{Thm_Bayes_Oracle_Risk}.
\end{proof}

The Bayesian False Discovery Rate (BFDR) was introduced by \citep{efron2002empirical} as
\begin{equation*}\textrm{BFDR}=P(H_{0}~\textrm{is true}\mid H_{0}~\textrm{is rejected})=\frac{(1-p)t_1}{(1-p)t_1+pt_2}.\end{equation*}
It has been seen that multiple testing procedures controlling BFDR at a small level $\alpha$ behave very well in terms of minimizing the misclassification error, see eg. \citep{bogdan2007empirical}.

Consider a fixed threshold rule based on $\tilde{S}_i$ with BFDR equal to $\alpha$. Under the mixture model (\ref{eqn_marginal}), a corresponding threshold value $c^2$ can be obtained by solving the equation
\begin{equation}\label{eqn_BFDR}
\frac{(1-p)e^{-c^2/2}}{(1-p)e^{-c^2/2}+pe^{-c^2/2u_t}}=\alpha
\end{equation}
\begin{thm}\label{Thm_BFDR} Consider a fixed threshold rule with BFDR=$\alpha=\alpha_t$. The rule is ABOS if and only if it satisfies the following two conditions\begin{equation}\label{eqn_cond1}r_{\alpha}/f\rightarrow 0, ~\textrm{where}~ r_\alpha=\alpha/(1-\alpha)\end{equation} and \begin{equation}\frac{2\log(r_{\alpha}/f)}{1-\frac{1}{u_t}}\rightarrow C.\label{eqn_cond2}\end{equation} The threshold for this rule is of the form \begin{equation}\label{eqn_Fixed}c_t^2=C-2\log(r_{\alpha}/f)+o(t).\end{equation}
\begin{proof}Suppose the test is ABOS.
Equation (\ref{eqn_BFDR}) is equivalent to  \begin{equation*}\frac{p}{1-p}\frac{\alpha}{1-\alpha}=e^{-c^2/2\left(1-u_t\right)}\end{equation*}
By theorem \ref{Thm_Si_Tilde_ABOS}, the right hand side goes to zero.  This implies left hand side = $r_{\alpha}/f$ goes to zero, establishing the first condition.\\
Simplifying Equation (\ref{eqn_BFDR}) and using $c_{it}^2u_t\rightarrow C$ we have,
\begin{equation}2\log(r_{\alpha}/f)=c_t^2+C+o(t)\label{eqn_Form}\end{equation}
that is, the threshold is of the form given by equation (\ref{eqn_Fixed}).\\
Furthermore, $c_{it}^2=C/u_t+o(t)$. Combining this with equation (\ref{eqn_Form}), we get the second condition.

Now we prove the converse. \\ Suppose a test with BFDR=$\alpha$ satisfies the two conditions. Let us define $z_t$ as $z_t=c_{it}^2u_t$. Such a test  satisfies equation (\ref{eqn_BFDR}). Hence,
\begin{equation*}2\log(r_{\alpha}/f)=z_t(1-\frac{1}{u_t}).\end{equation*} Combining this with equation (\ref{eqn_cond2}), we have $z_t\rightarrow C$. \\Also, from equation (\ref{eqn_Fixed}), $2\log(r_{\alpha}/f)=z_t-c_t^2$. Combining this with equation (\ref{eqn_cond1}) and $z_t\rightarrow C$, we have $c_t^2\rightarrow\infty$.\\ Now by using theorem \ref{Thm_Si_Tilde_ABOS}, the test is ABOS.\qed
\end{proof}
\end{thm}

The cut-off value $c$ for $\tilde{S}_i$ is a solution of equation (\ref{eqn_BFDR}). This $c$ is still a function of unknown parameters and cannot be used in practice. Note that when the sparsity parameter $p$ and model parameters $\lambda_j$ are fixed, $c$ is a monotone decreasing function of the level $\alpha$ of the test. Hence for implementation purpose, instead of trying to explicitly obtaining $c$, we fix the proportion of tests for which we'll reject the null hypothesis. This amounts to fixing the proportion  of assets to include in the portfolio. This gives control over the size of the resulting portfolio. This is also desirable from the financial point of view as competing portfolios are comparable only when they are of the same size.

\section{Simulation Study}\label{Section_Simulation}
In this section, we present four different simulation studies. In experiment 1, we compare the performance of $S_i$ and $\tilde{S}_i$. In the second experiment, we compare the performance of the Bayes oracle estimator $\tilde{S}_i$ with the other methods like diffuse prior and LARS-LASSO \cite{Fan.Zhang.Yu.2012}. In the third experiment, we compare the probability that the portfolio selected by the proposed $\tilde{S}_i$ will contain the  oracle set as a function of market size, sample size, and idiosyncratic risk. In the fourth experiment, we study the effect of the number, $k$, of factors on the performance of $\tilde{S}_i$.

For the first three experiments, we consider the one factor CAPM to keep things simple and simulate the data from a true model given by equation (\ref{eqn_CAPM}) with $\sigma_i=\sigma$ for all $i$.  Without loss of generality, we assume that the first $[pP]$ many stocks are not fairly priced. That is, we simulate $\alpha_i$ and $\beta_i$ for those stocks from $N(0,0.1)$ and $N(1,0.1)$ respectively. For the rest of the stocks $\alpha$ and $\beta$ are set to (0,1).

\vspace{2mm}

\begin{figure}[ht]
  \centering
  \caption{Performance of $S_i$ and $\tilde{S}_i$ in Experiment 1. We simulate 1000 datasets. In each dataset, we consider the 100 stocks, i.e., $P=100$ with 20 and 50 days of data, i.e., sample size $n=20$ and $n=50$. We allow the sparsity parameter $p$ to vary from 0.01 to 0.9 by an interval of 0.01. We choose two different values of $\sigma$ as $0.1$ and $0.05$. Also, $\Lambda_0$ is defined in Experiment 1.  For each dataset, we compute $S_i$ and $\tilde{S}_i$ and make a decision. Based on the decision over 1000 datasets we compute type-I error, type-II error, Bayesian False Discovery Rate (BFDR), and the probability of misclassification (PMC). Overlapping performance of $S_i$ and $\tilde{S}_i$ indicates that both statistics are equivalent.}
 \begin{tabular}{cc}\\
\includegraphics[width=.45\linewidth]{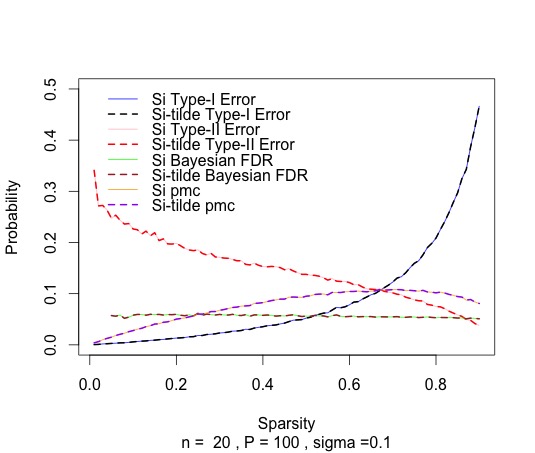}
&
\includegraphics[width=.45\linewidth]{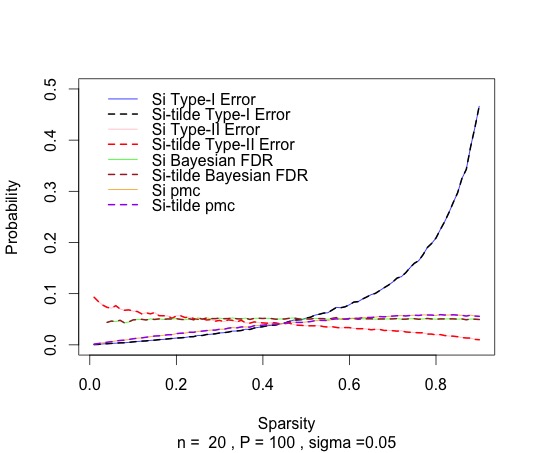}\\
$n=20$,$\sigma=0.1$ & $n=20$,$\sigma=0.05$\\
\includegraphics[width=.45\linewidth]{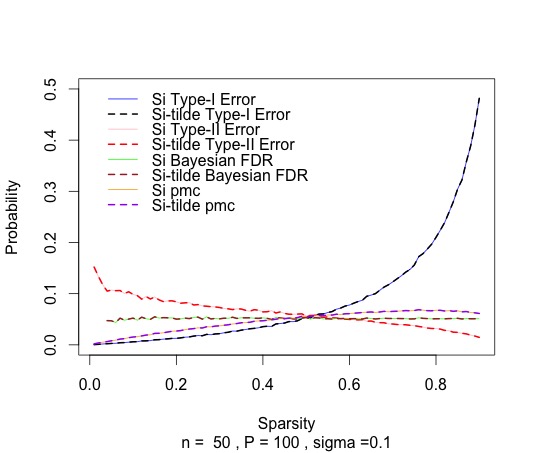}
&
\includegraphics[width=.45\linewidth]{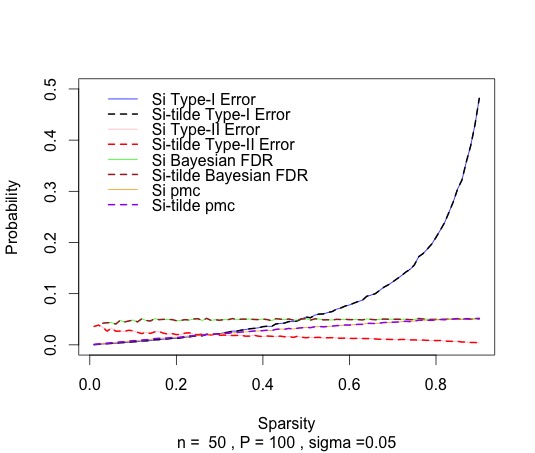}\\
$n=50$,$\sigma=0.1$ & $n=50$,$\sigma=0.05$
\end{tabular}
\label{Fig_sim_study1}
\end{figure}

\begin{figure}[ht]
  \centering
\caption{Here we study the effect of increasing $P$ on $S_i$ and $\tilde{S}_i$ in Experiment 1.  We simulate 1000 datasets.  In each datasets,  we consider 10 and 500 stocks, i.e., $P=10$ and $P=500$ with 20 days of data, i.e., sample size $n=20$. We consider $\sigma=0.1$.  We allow the sparsity parameter $p$ to vary from 0.01 to 0.9 by an interval of $0.01$. For each dataset, we compute $S_i$ and $\tilde{S}_i$ and make a decision. Based on the decision over 1000 datasets we compute type-I error, type-II error, Bayesian False Discovery Rate (BFDR), and the probability of misclassification (PMC). }
 \begin{tabular}{cc}\\
\includegraphics[width=.45\linewidth]{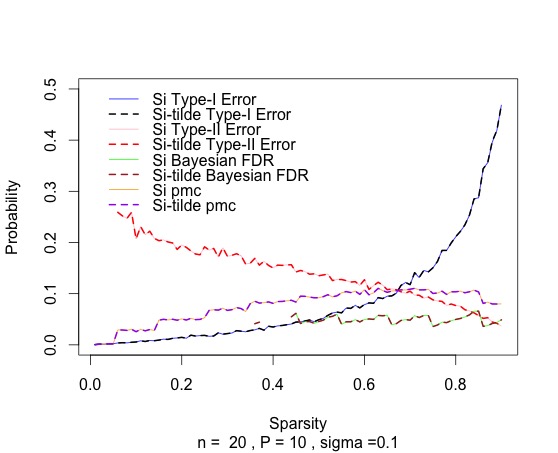}
&
\includegraphics[width=.45\linewidth]{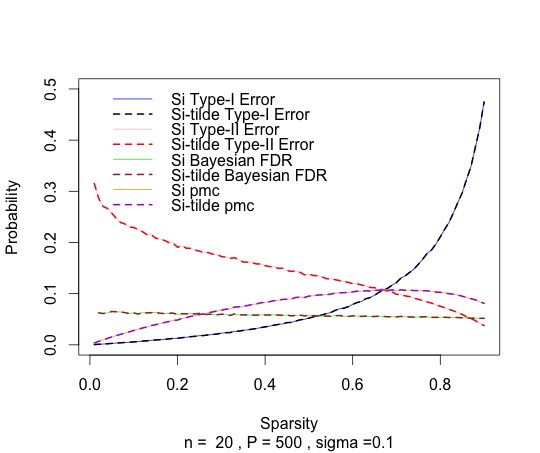}\\
(a) $P=10$, $n=20$, $\sigma=0.1$ & (b) $P=500$, $n=20$, $\sigma=0.1$\\
\end{tabular}
  \label{Fig_sim_study1_fig2}
\end{figure}

\noindent \textbf{Experiment 1}:  In this study, we consider two different choices of $\tilde{P}$, i.e., $\tilde{P}=100$ and $500$ and the sample size is varied from $n=20$ to $n=50$ by an interval of $5$. Note that due to space constraint we present the result for $n=20$ and $n=50$ in figure \ref{Fig_sim_study1}. We allow the sparsity parameter $p$ to vary from 0.01 to 0.9 by an interval of 0.01. We choose two different values of $\sigma$ as $0.1$ and $0.05$. Finally, we set
$\Lambda_0=\left(\begin{matrix}
0.5 & 0.3\\
0.3 & 0.7
\end{matrix}\right)$. For all these different choices of $n$, $\tilde{P}$, $\alpha$, $\beta$, $\sigma$; we simulate 1000 datasets. For each dataset, we compute $S_i$ and $\tilde{S}_i$ for all $i=1,2,\cdots,P$. In standard Bayesian variable selection strategy, either posterior inclusion probabilities, see \cite{Datta.Ghosh.2013}, or credible intervals, see \cite{vanderPas2017} are used. We adopt an alternate strategy of selecting a fixed number of assets for the final portfolio. For example, in a given simulated dataset, we calculated $\tilde{S}_i$ statistics for the $i^{th}$ asset for all $i=1,2,...\tilde{P}$. Then we select $P=25$ many assets for which the corresponding $\tilde{S}_i$  are highest. The reason we avoid the standard Bayesian variable selection strategy, is as follows. The fixed threshold strategy or probability inclusion strategy may select $P_1$ assets in one dataset. However, in another datasets, it may select $P_2$ assets. Then these two portfolios are not comparable, as the covariance matrix are of different dimensions. We want to keep the portfolio size to be the same in all simulated 1000 datasets for fair comparison. We follow the same portfolio selection strategy proposed here, in all four experiments.  Based on the decision over 1000 datasets we compute type-I error, type-II error, Bayesian False Discovery Rate (BFDR), and the probability of misclassification (PMC); and present the results in figures \ref{Fig_sim_study1} and \ref{Fig_sim_study1_fig2}. We report the following observations for Experiment 1.

\noindent \textbf{Observations}
\begin{enumerate}
\item As the sparsity tends to 0, the type-I error goes to 0 in all four panels of figure \ref{Fig_sim_study1}. In all panels of figure \ref{Fig_sim_study1}, sparsity is close to 0 and the type-II error does not shoot to 1. The bounded type-II error guarantees particular statistical power of the ABOS. This observation verifies Theorem \ref{Thm_T1_error_goes_to_0}.
\item From all four panels of figure \ref{Fig_sim_study1}, we observe, for different choices of $n$, $P$, $\sigma$ and sparsity $p$, all the four metrics of $S_i$ and $\tilde{S}_i$ overlap. Hence this verifies Theorem \ref{Thm_Si_Tilde_ABOS}.
\item As the sample size $n$ increases, both the probability of type-II error and the probability of misclassification drop. It indicates increasing statistical power even when the sparsity parameter $p$ is near zero.
\item As $p\rightarrow 1$, i.e., the model becomes dense, one should not use this test, as type-I error increases. However, up to 0.5 of the sparsity, the type-I error stays below $5\%$ level.
\item In all four panels of figure \ref{Fig_sim_study1} the BFDR is about 0.05 irrespective the value of $n$, $P$, $\sigma$ and $p$.
\item Figure \ref{Fig_sim_study1_fig2} indicates that by increasing the number of stocks from $P=10$ to $P=500$, all the metrics of the test become smoother.
\end{enumerate}

\vspace{5mm}

\begin{figure}[ht]
\centering
\caption{Here we present the performance comparison of ABOS, Diffuse prior and LARS-LASSO from Experiment 2. We simulate 1000 data independently, each with 500 stock and 20 days of return. We consider $\sigma=0.1$ and $\Lambda_0$ defined in the Experiment 2. Based on the decision on 1000 datasets we compute the probability of Type-I error, Type-II error, Bayesian False Discover Rate (BFDR) and the probability of misclassification (PMC). Here we present the results. As the sparsity tends to 0, the type-I error of ABOS goes to 0. In likelihood testing (i.e., with diffuse prior), the type-I error is fixed at $5\%$ level throughout the different values of sparsity. The LARS-LASSO method also demonstrates a flat behavior.  If we compare the type-II error, BFDR, and the PMC for all three methods, the ABOS test proposed in this paper is uniformly better than the other two methods. It means when the market is nearly efficient with the $\alpha$ od a few stocks being non-zero, the ABOS should be the preferred choice.}
\includegraphics[width=1\linewidth]{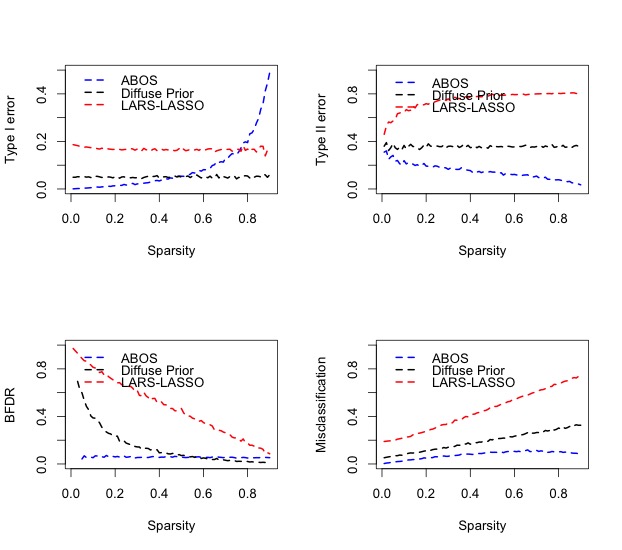}
\label{Fig_sim_study2}
\end{figure}

\noindent \textbf{Experiment 2}: In this study, we consider $P=500$, the sample size $n=20$, $\sigma=0.1$ and
$\Lambda_0=\left(\begin{matrix}
0.5 & 0.3\\
0.3 & 0.7
\end{matrix}\right)$. For all these choices of parameters, we simulate 1000 datasets. For each dataset, we compute $\tilde{S}_i$ and make a decision. We also make the decision using diffuse-prior and LARS-LASSO technique and compare against the oracle. Based on the decision in 1000 datasets we compute type-I error, type-II error, Bayesian False Discover Rate (BFDR) and the probability of misclassification (PMC) and present the results in figure \ref{Fig_sim_study2}.

\noindent \textbf{Observations}
\begin{enumerate}
\item As the sparsity tends to 0, the type-I error of ABOS goes to 0. In likelihood testing, the type-I error is fixed at $5\%$ level throughout the different values of sparsity. The probability of type-I error for the LARS-LASSO method also demonstrates a flat behavior, like the diffuse prior. However, it is more than that of the diffuse prior method. Similarly, the probability of type II error, BFDR, and the probability of misclassifications for LARS-LASSO are uniformly higher than the diffuse prior method.
\item If we compare the type-II error, BFDR, and the PMC for all three methods, the ABOS test proposed in this paper is uniformly better than the other two methods.
\end{enumerate}
\noindent \textbf{Note}: We tried to compare the ABOS test against Hierarchical Bayes (HB) with horseshoe prior method. However, given the computational power, it took about three days to implement the HB method for 1000 simulated datasets, for one fixed sparsity parameter. For each dataset, we simulated 25000 MCMC simulations after 5000 burn-in. For experiment 2, we consider the sparsity ranges from 0.01 to 0.9 by an interval of 0.01. For one sparsity value it was taking approximately three days, and for all 90 possible values, it will take about 270 days, assuming no possible disruption in the systems. Hence we could not implement the comparison due to the lack of computational resources. Therefore we leave this task as a future research project.

\vspace{5mm}

\noindent \textbf{Experiment 3}:  The objective of this study is to compare the portfolio return from true oracle portfolio and portfolio selected via the ABOS test method. We simulate 1000 datasets. In each dataset, we simulate 40 samples from the model described in experiment 1. We consider 20 samples for training and the rest for testing. Throughout we assume that the market size is $P=500$.  We consider the portfolio size to be $\tilde{P}=100$ and $\tilde{P}=50$. We consider two different choices of $\sigma$, 0.03 and 0.01 respectively. We assume $5\%$ (i.e., $q=25$) of the stocks have non-zero $\alpha$. We consider the oracle portfolio, (denoted as $A_q^{\tilde{P}}$) and the ABOS portfolio (denoted as $B^{\tilde{P}}$) for the study. In the oracle portfolio, $25$ stocks will always be selected along with 75 other randomly selected stocks. In the ABOS portfolio, all stocks were selected based on $\tilde{S}_i$. We consider equal weight for both portfolios.

\begin{figure}[ht]
  \centering
  \caption{ Side-by-side box plot of the probability that the selected portfolio contains the oracle portfolio. Boxplot of 1000 out of sample returns of the true oracle portfolio and the portfolio selected based on $\tilde{S}_i$. Visual inspection tells us that the performance of the ABOS portfolio is similar/equivalent to the oracle portfolio.}
\begin{tabular}{cc}
\includegraphics[width=.4\linewidth]{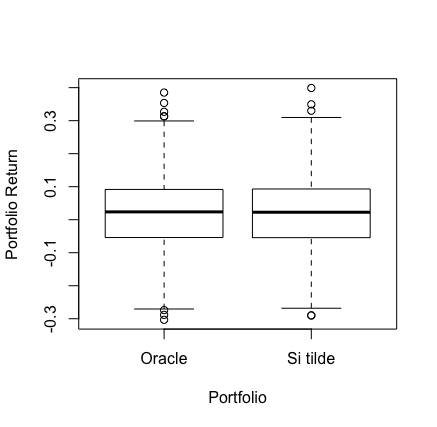}
& \includegraphics[width=.4\linewidth]{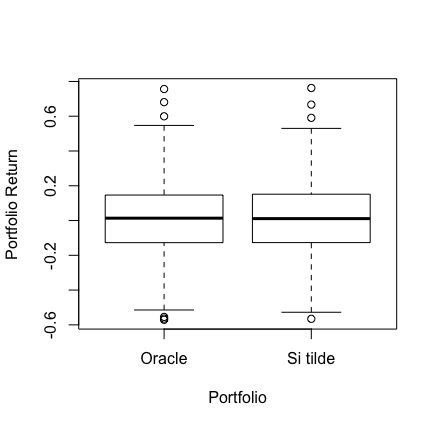}\\
$\tilde{P}=100$, $\sigma=0.03$
&$\tilde{P}=50$, $\sigma=0.03$\\
\includegraphics[width=.4\linewidth]{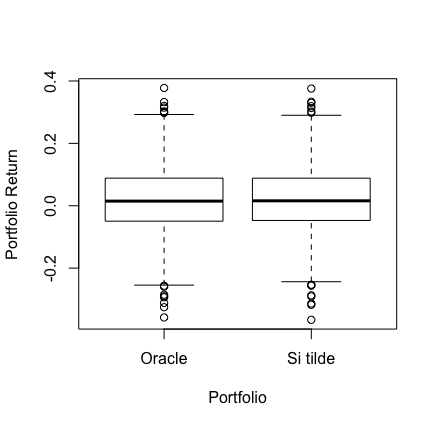}
&\includegraphics[width=.4\linewidth]{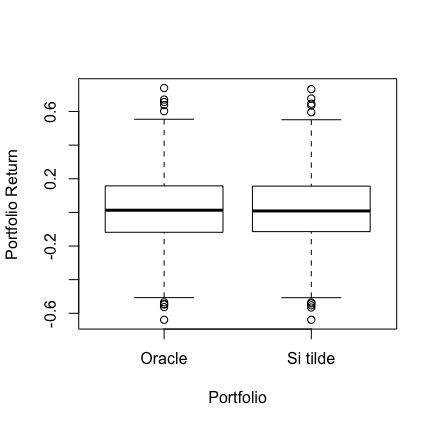}\\
$\tilde{P}=100$, $\sigma=0.01$
&$\tilde{P}=50$, $\sigma=0.01$
\end{tabular}
  \label{Fig_sim_study3}
\end{figure}

\noindent \textbf{Observations}:
\begin{enumerate}
\item In table (\ref{tbl_return}), we present the out sample median return from 1000 synthetic datasets. The median return for both the true oracle portfolio and the ABOS portfolio are similar for four different choices of $\tilde{P}$ and $\sigma$.
\item In figure (\ref{Fig_sim_study3}), we present side-by-side boxplot of returns from 1000 synthetic datasets for the true oracle portfolio and the ABOS portfolio. Visual inspection tells us the performance of the ABOS portfolio is similar/equivalent to the oracle portfolio.
\end{enumerate}

\vspace{5mm}

\begin{figure}[ht]
  \centering
  \caption{The line-plot of the probability, that the selected portfolio contains the oracle portfolio. The probability is estimated using 1000 out-sample returns. On the x-axis, we consider varying idiosyncratic risks. The statistical power of the ABOS portfolio is uniformly better for the four-factor model than the one factor CAPM.}
\begin{tabular}{cc}
\includegraphics[height=6cm,width=6cm]{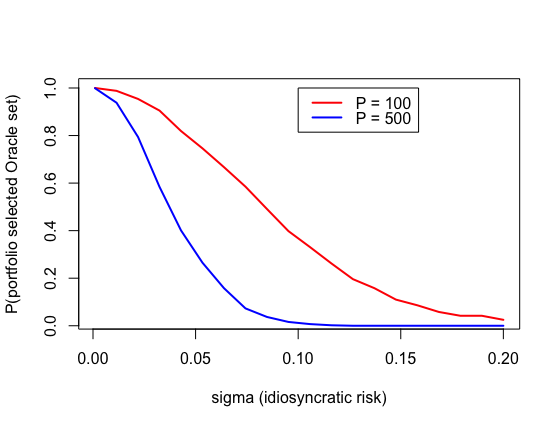}
& \includegraphics[height=6cm,width=6cm]{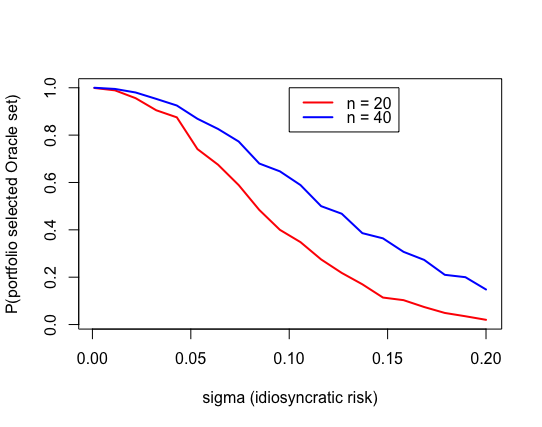}\\
One factor CAPM with $\tilde{P}=100$ vs $500$
&One factor CAPM with $n=20$ vs $40$ \\
and $n=20$ in both case & and $P=100$ in both cases \\ \hline
\includegraphics[height=6cm,width=6cm]{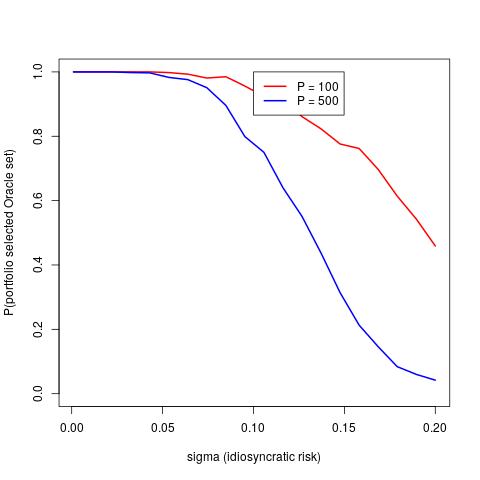}
&\includegraphics[height=6cm,width=6cm]{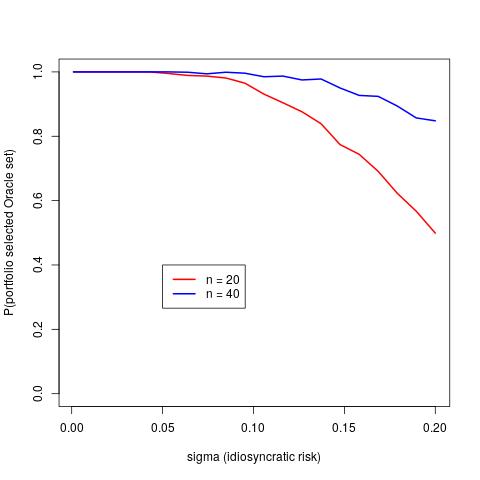}\\
The 4 factor model $\tilde{P}=100$ vs $500$
& The 4 factor model $n=20$ vs $40$\\
and $n=20$ in both case & and $P=100$ in both cases \\
\end{tabular}
  \label{Fig_sim_study4}
\end{figure}

\noindent \textbf{Experiment 4}: The objective of this study is to investigate the effect of the number of factors on the performance of the portfolio selection procedure.  We compare the probability that the selected portfolio contains the oracle portfolio, between the CAPM and the four-factor model. Note that the CAPM model is effectively a one-factor model, where risk-premium due to market index is the only factor.
\noindent \textbf{Observations}:
\begin{enumerate}
\item In figure (\ref{Fig_sim_study4}), we present the line-plot of the probability that the selected portfolio contains the oracle portfolio. Visual inspection indicates that the statistical power of the ABOS portfolio is uniformly better for the four-factor model than the one factor CAPM.
\end{enumerate}
\section{Empirical Study}\label{Section_Empirical}

In the empirical study, we considered about thirteen years of daily returns from Jan 2006 to Oct 2018. The purpose of choosing this period is to study the behavior of the methods especially during the stress period of 2008 and 2011. We considered about 500 stocks listed with the New York Stock Exchange. We downloaded the daily adjusted close prices of the 500 stocks on the 1st of November, 2018.  Also, we downloaded the daily closing value of S\&P 500 index, Dow Jones Industrial Average (DJI), NYSE Composite (NYSEC), Russell 2000 index and CBOE Volatility Index (aka. VIX). All data was downloaded from Yahoo Finance (\url{https://finance.yahoo.com/}). All the \texttt{R} code and data are available in the following GitHub repository: \url{https://github.com/sourish-cmi/sparse_portfolio_Bayes_multiple_test}. We considered S\&P 500 index as the benchmark market index and indices like DJI, NYSEC, Russell 2000 and VIX as additional factors in a five-factor model. Note that during parts of this thirteen year period certain stocks were not available. Initial years daily closing prices of about 400 stocks were available. We implemented our analysis with available stock prices.

We implemented the following monthly analysis. On the $t^{th}$ month, we run the modeling procedure described in Section (\ref{Section_Methodology}) over the daily return of the $t^{th}$ month. There are $P=500$ many assets with excess returns over risk-free rate $r_1,...,r_P$ in the market and $P \gg n$. Here $n$ is typically 22 or 23 days of return, as there are only that many business days in a month. We select $\tilde{P}=25$ many assets using the methodology described in Section \ref{Section_Methodology} applied to the daily return of $t^{th}$ month. We select the stocks which are under-priced and our revised portfolio for $(t+1)^{st}$ month would be composed of these under-priced stocks. Once we select $\tilde{P}=25$ many assets for the portfolio, then the problem reduces to portfolio allocation. Typically, portfolio managers resort to the solution from the Markowitz's optimization. However, we realize if we resort to the Markowitz's optimization, we will not be able to distinguish, if the enhanced performance is due to portfolio selection or due to Markowitz's portfolio allocation. To nullify the effect of the portfolio allocation, we considered the equal-weighted portfolio throughout. Note that we also separately implemented the results with Markowitz's optimization and presented in Appendix B in the supplementary materials. We observe out of sample risk and return are different for Markowitz's weighted portfolio. But we do not observe any major significant improvement due to Markowitz's optimization.

\begin{figure}[ht]
  \centering
  \caption{The line-plot of out of sample annualised return, annualised risk-adjusted return, Value at Risk and Volatility for S\&P 500, CAPM, LARS-LASSO, Factor Model with horseshoe prior, Factor model with ABOS for equal weight portfolio  in empirical study.}
\begin{tabular}{cc}
\includegraphics[height=6cm,width=6cm]{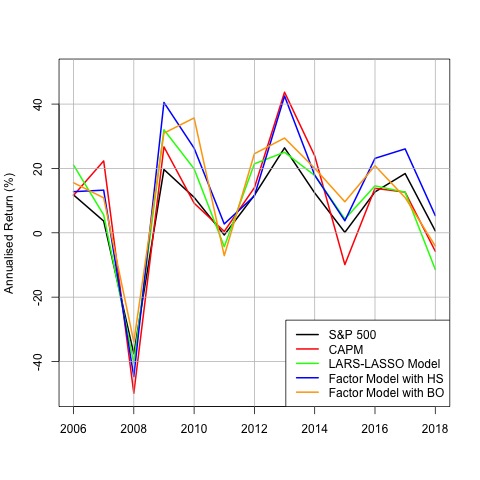}
& \includegraphics[height=6cm,width=6cm]{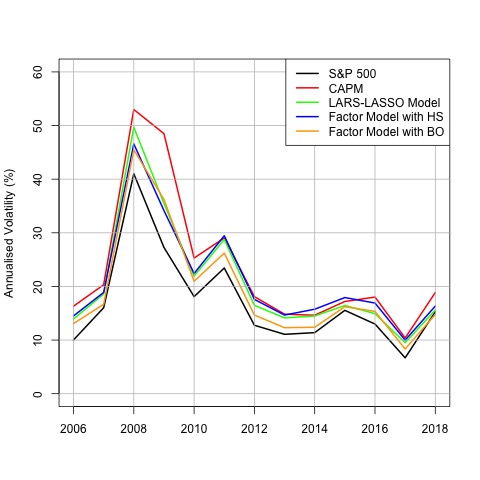}\\
(a) Annualised Return (\%)
& (b) Volatility\\
\hline
\includegraphics[height=6cm,width=6cm]{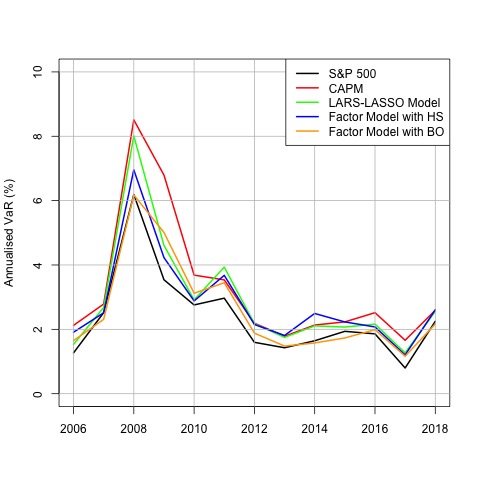}
&\includegraphics[height=6cm,width=6cm]{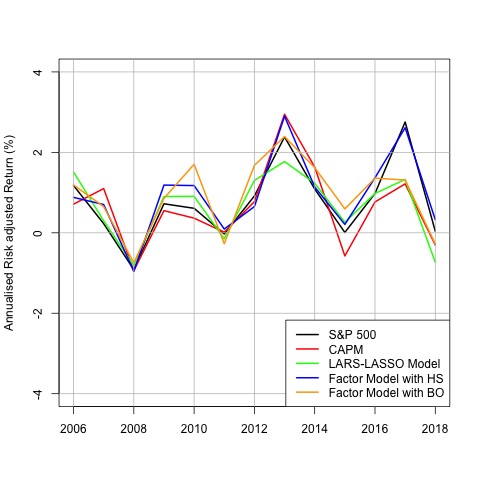}\\
(c) Value at Risk
& (d) Annualised  Risk-adjusted Return \\
\end{tabular}
  \label{Fig_portfolio_anualised_sum_equal_wt}
\end{figure}

We invest in the selected stocks for the month $t+1$ and calculate the out-sample return of the portfolio and S\&P 500 Index at the same time. We considered the out-sample period to be from Jan 01, 2006 to Oct 31, 2018. For example, we run the statistical processes on the stock price of Dec 2007 and identify the 25 stocks and construct the portfolio using the equal weights for Jan 2008 and invest only in these stocks. We repeat the process for each month. The performance for portfolio with equal weight is presented in figure \ref{Fig_portfolio_anualised_sum_equal_wt}.

\begin{figure}[ht]
\centering
\caption{Performance of a different portfolios in out-sample}
\label{Fig_portfolio_Value}
\includegraphics[ height=18cm,width=15cm]{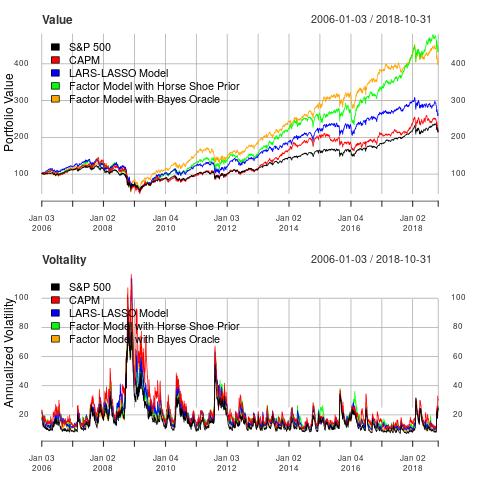}
\end{figure}

Figure \ref{Fig_portfolio_Value}, shows the performance of four different strategies in out-sample. Here S\&P 500 index is being considered as benchmark portfolio, as many investors invest in S\&P500 index fund as passive investors. If we look at the figure (\ref{Fig_portfolio_anualised_sum_equal_wt}:a), except for two years (2008,2012) the `factor models with horseshoe' selection strategy outperforms the benchmark index in terms of annual return. Out of 13 years, there are five years (2008, 2011, 2013, 2017, 2018) where Fan's LARS-LASSO selection process under-performs compared to benchmark in terms of annual return. Similarly, out of the same 13 years, there are six years (2006, 2008, 2010, 2015, 2017, 2018) the CAPM selection process under performs compared to benchmark in terms of annual return. Except for three years  (2011, 2017, 2018), the factor models with Bayes oracle strategy outperforms the benchmark index fund in terms of annual return. In terms of annual return all four portfolio selection processes outperform the benchmark in all the thirteen years, but we cannot clearly say one strategy outperforms the other for all years.

In figure (\ref{Fig_portfolio_anualised_sum_equal_wt}:b) we present the out-sample annualized volatility of all the selection strategies and in figure \ref{Fig_portfolio_Value} we present the daily annualized volatility of all four selection strategies in out-of-the sample return. The volatility is estimated with the GARCH(1,1) model. The volatility risk for all four strategies mimic the volatility risk of the benchmark S\&P 500 index and the volatility risk for all the four strategies is always in the neighborhood of the S\&P 500 index. However careful visual inspection indicates that all four strategies have systematically marginally higher volatility risk compared to the S\&P 500 index. The volatility risk of the Bayes oracle strategy is always consistently lower among the four strategies. Figure (\ref{Fig_portfolio_anualised_sum_equal_wt}:c) presents the `Value at Risk' (VaR) of the different strategies in out-sample return. The VaR of the Bayes oracle strategy tends to be lower among the four strategies.

Figure (\ref{Fig_portfolio_anualised_sum_equal_wt}:d) presents the `risk-adjusted return' of different strategies in out-sample. There is no single consistent winner. There are five years where hierarchical Bayes strategy has the highest risk-adjusted return. For all the years except 2017, either of the four strategies has a higher risk-adjusted return compared to the benchmark S\&P 500 index.  The risk-adjusted return indicates the possible existence of inefficiency in the market. \emph{There is no clear winning strategy in terms of annual return. However all four strategies indicate the possible existence of inefficiency in the market}.

We implemented the Markowitz's portfolio optimization technique instead of equal portfolio allocation and added the results in the Appendix B. We observe that the out of sample risk and return are different for the Markowitz's weighted portfolio. But we do not observe any significant improvement due to Markowitz's optimization.

\section{Discussion}\label{Section_Discussion}

We presented Bayesian portfolio selection strategy, via the $k$ factor model. If the market is information efficient, the proposed strategy will mimic the market; otherwise, the strategy will outperform the market. The strategy depends on the selection of a portfolio via Bayesian multiple testing methodologies for the parameters of the model. We present the ``discrete-mixture prior" model and ``hierarchical Bayes model with horseshoe prior."  We prove that under the asymptotic framework of \cite{Datta.Ghosh.2013}; the Bayes rule attains the risk of Bayes oracle up to $O(1)$ with a  constant close to the constant in the oracle. The Bayes oracle test guarantees statistical power by providing the upper bound of the type-II error.

A simulation study indicates that Bayes oracle test has an increasing statistical power with increasing sample size when the sparsity parameter $p$ is near zero. As $p \rightarrow 1$, i.e., the model becomes dense, one should not use the Bayes oracle test. However, up to $p=0.5$ of the sparsity parameter, the type-I error stays below 5\% level. Hence, we can use the test even when $p=0.5$, i.e., the model is moderately sparse. It means the proposed Bayes oracle test is suitable for the efficient market with few stocks inefficiently priced. The statistical power of the Bayes oracle portfolio is uniformly better for the $k$-factor model ($k>1$) than the one factor CAPM.

We present an empirical study, where we considered 500 stocks from the New York Stock Exchange (NYSE), and S\&P 500 index as the benchmark, over the period from the year 2006 to 2018. We presented the out-sample risk and return performance of the four different strategies and compared with the S\&P 500 index. Empirical results indicate the existence of inefficiency in the market, and it is possible to propose a strategy which can outperform the market.

In the literature on `horseshoe’ priors, both the local shrinkage and the global shrinkage parameters have a heavy-tailed half-Cauchy prior.
Although that is theoretically possible for our analysis, we have chosen to specify a conjugate Wishart prior on the precision matrix $\Lambda$ that contains the local shrinkage parameters and a half-Cauchy prior on the global shrinkage parameter $\tau$. This is motivated by concerns of computational feasibility. We are addressing an applied problem in finance where the procedure has to be implemented in an extremely high dimensional set-up. In our case, it is monthly returns for 10 years on 500 stocks. The conjugacy of the Wishart aids the computation immensely.

\bibliographystyle{plainnat}
\bibliography{Biblio-Database}

\appendix
\section{Appendix }
Markowitz portfolio optimization can be expressed as the following quadratic programming problem:
\begin{eqnarray}\label{eqn_optim}
\mathrm{min} ~~ \mathbf{w}'\Sigma \mathbf{w}\quad \text{subject to}\quad\mathbf{w}'\mathbf{1}_P=1
~~\mathrm{ and}\quad \mathbf{w}'{\boldsymbol{\mu}}= \mu_k.
\end{eqnarray}
Here $\mathbf{1}_P$ is a $P$-dimensional vector with one in every entry and $\mu_k$ is the desired level of return.

The portfolio covariance can be decomposed into two parts as,
\begin{eqnarray*}
\mathbf{w}'\Sigma \mathbf{w} &= &\mathbf{w}'[\bm{B}^T\bfSigma_m\bm{B}+\bfSigma_{\epsilon}] \mathbf{w}\\
& = & \mathbf{w}'\bm{B}^T\bfSigma_m\bm{B}\mathbf{w} + \mathbf{w}' \bfSigma_{\epsilon} \mathbf{w},
\end{eqnarray*}
where first part  explains the portfolio volatility due to market volatility and the second part explains portfolio volatility due to idiosyncratic behaviour of the stock. We assume $\sigma_i$'s are bounded $\forall i$. Then
\begin{eqnarray*}
\mathbf{w}' \bfSigma_{\epsilon} \mathbf{w} &=& \sum_{i=1}^P\omega_i^2\sigma_i^2\\
&\leq& \sigma_{max}^2\sum_{i=1}^P\omega_i^2,~~\sigma_{max}^2=\max\{\bfSigma_{\epsilon}\}<\infty,\\
&\leq& \sigma_{max}^2M_{\omega:P}\sum_{i=1}^P\omega_i,~~M_{\omega:P}=\max\{\mathbf{w}\},\\
&=& \sigma_{max}^2M_{\omega:P}.
\end{eqnarray*}
Clearly, if $P\rightarrow \infty$ and $M_{\omega:P} \rightarrow 0 \implies \mathbf{w}' \bfSigma_{\epsilon} \mathbf{w} \rightarrow 0$. Hence we have the following result.

\begin{result}\label{result_idiosyn_wash_out}
Under the CAPM model (\ref{eqn_CAPM}), covariance matrix (\ref{eqn_cov_matrix}) and assumption (\ref{eqn_optim}),
if $P\rightarrow \infty$ and $M_{\omega:P} \rightarrow 0$ and $\sigma_{max}^2=\max\{\bfSigma_{\epsilon}\}<\infty$, then
$$
\lim_{P\rightarrow \infty} \mathbf{w}' \bfSigma_{\epsilon} \mathbf{w} =0.
$$
\end{result}

\begin{remark}
Under the standard asset pricing theory, as the size of the portfolio increases, and the maximum weight of any asset is bounded, the idiosyncratic risk of the portfolio will be washed out. The portfolio's risk and return will be a function of the systematic risk explained by major market indices.
\end{remark}

\clearpage
\section{Appendix }

\begin{table}[ht]
\centering
\caption{Median return (of 1000 out of sample datasets) of the true oracle portfolio and portfolio selected by the proposed $\tilde{S}_i$ method. It indicates ABOS selection and oracle selection are equivalent.}\label{tbl_return}
\begin{tabular}{rrrr}
  \hline
&&&\\
 $\tilde{P}$ & $\sigma$ & Oracle Portfolio Return & $\tilde{S}_i$ Portfolio Return \\
  \hline
100 & 0.03 & 0.024 & 0.023 \\
50   & 0.03 & 0.014 & 0.013 \\
100 & 0.01 & 0.015 & 0.016 \\
50   & 0.01 & 0.013 & 0.010 \\
   \hline
\end{tabular}
\end{table}

\begin{table}[ht]
\centering
\caption{Out-sample annual return of equal weight portfolio. The blue value indicates the maximum return compared to other strategies in a particular year.}\label{tbl_return_equal_wt}
\begin{tabular}{rrrrrr}
  \hline
 & S\&P 500 & CAPM & LARS-LASSO Model & Factor Model with HS & Factor Model with BO \\
  \hline
2006 & 11.78 & 11.62 & \textcolor{blue}{21.06} & 12.78 & 15.56 \\
2007 & 3.65 & \textcolor{blue}{22.33} & 5.56 & 13.26 & 10.87 \\
2008 & -37.58 & -49.85 & -40.44 & -44.70 & \textcolor{blue}{-33.89} \\
2009 & 19.67 & 26.70 & 32.02 & \textcolor{blue}{40.51} & 30.97 \\
2010 & 11.02 & 9.21 & 19.89 & 26.31 & \textcolor{blue}{35.70} \\
2011 & -0.68 & 0.43 & -4.32 & \textcolor{blue}{2.75} & -7.12 \\
2012 & 11.68 & 14.03 & 21.44 & 11.46 & \textcolor{blue}{24.56} \\
2013 & 26.39 & \textcolor{blue}{43.69} & 25.04 & 42.50 & 29.45 \\
2014 & 12.39 & \textcolor{blue}{23.93} & 17.83 & 17.92 & 20.06 \\
2015 & 0.14 & -9.89 & 4.28 & 3.73 & \textcolor{blue}{9.64} \\
2016 & 12.70 & 13.82 & 14.58 & \textcolor{blue}{23.14} & 20.85 \\
2017 & 18.42 & 12.63 & 12.61 & \textcolor{blue}{26.07} & 10.93 \\
2018 & 0.59 & -5.73 & -11.39 & \textcolor{blue}{5.34} & -4.11 \\
   \hline
   \hline
\end{tabular}
\end{table}

\begin{table}[ht]
\centering
\caption{Annualized Volatility of out-sample return of equal weight portfolio.}\label{tbl_vol}
\begin{tabular}{rrrrrr}
  \hline
& S\&P 500 & CAPM & LARS-LASSO Model & Factor Model with HS & Factor Model with BO \\
  \hline
2006 & 10.02 & 16.32 & 13.93 & 14.50 & 13.08 \\
2007 & 16.02 & 20.31 & 18.59 & 18.88 & 16.67 \\
2008 & 41.02 & 52.99 & 49.68 & 46.57 & 45.47 \\
2009 & 27.27 & 48.49 & 35.50 & 34.14 & 36.22 \\
2010 & 18.10 & 25.29 & 21.93 & 22.41 & 20.96 \\
2011 & 23.44 & 29.04 & 28.76 & 29.45 & 26.22 \\
2012 & 12.76 & 18.10 & 16.48 & 17.58 & 14.66 \\
2013 & 11.07 & 14.83 & 14.14 & 14.66 & 12.31 \\
2014 & 11.38 & 14.63 & 14.49 & 15.75 & 12.39 \\
2015 & 15.54 & 17.19 & 16.45 & 17.93 & 16.22 \\
2016 & 12.99 & 18.02 & 14.93 & 16.90 & 15.35 \\
2017 & 6.69 & 10.39 & 9.50 & 9.98 & 8.34 \\
2018 & 15.26 & 18.85 & 15.59 & 16.33 & 14.78 \\
   \hline
\end{tabular}
\end{table}

\begin{table}[ht]
\centering
\caption{Value at Risk (VaR) of out-sample return of equal weight portfolio.}\label{tbl_VaR}
\begin{tabular}{rrrrrr}
  \hline
& S\&P 500 & CAPM &  LARS-LASSO Model & Factor Model with HS & Factor Model with BO \\
  \hline
2006 & 1.27 & 2.12 & 1.53 & 1.91 & 1.65 \\
2007 & 2.52 & 2.79 & 2.68 & 2.51 & 2.31 \\
2008 & 6.19 & 8.51 & 8.01 & 6.96 & 6.18 \\
2009 & 3.54 & 6.79 & 4.61 & 4.23 & 5.01 \\
2010 & 2.76 & 3.69 & 2.91 & 2.88 & 3.12 \\
2011 & 2.97 & 3.54 & 3.93 & 3.68 & 3.45 \\
2012 & 1.60 & 2.19 & 2.18 & 2.15 & 1.88 \\
2013 & 1.43 & 1.78 & 1.75 & 1.81 & 1.48 \\
2014 & 1.65 & 2.13 & 2.11 & 2.49 & 1.58 \\
2015 & 1.94 & 2.23 & 2.07 & 2.23 & 1.73 \\
2016 & 1.86 & 2.52 & 2.16 & 2.07 & 1.99 \\
2017 & 0.80 & 1.66 & 1.27 & 1.20 & 1.16 \\
2018 & 2.25 & 2.59 & 2.56 & 2.61 & 2.16 \\
 \hline
\end{tabular}
\end{table}

\begin{table}[ht]
\centering
\caption{Risk adjusted out-sample return of equal weight portfolio.  The blue value indicates the maximum risk-adjusted return compared to other strategies in a particular year.}\label{tbl_sharpe}
\begin{tabular}{rrrrrr}
  \hline
& S\&P 500 & CAPM &  LARS-LASSO Model & Factor Model with HS & Factor Model with BO \\
  \hline
2006 & 1.18 & 0.71 & \textcolor{blue}{1.51} & 0.88 & 1.19 \\
2007 & 0.23 & \textcolor{blue}{1.10} & 0.30 & 0.70 & 0.65 \\
2008 & -0.92 & -0.94 & -0.81 & -0.96 & \textcolor{blue}{-0.75} \\
2009 & 0.72 & 0.55 & 0.90 & \textcolor{blue}{1.19} & 0.85 \\
2010 & 0.61 & 0.36 & 0.91 & 1.17 & \textcolor{blue}{1.70} \\
2011 & -0.03 & 0.01 & -0.15 & \textcolor{blue}{0.09} & -0.27 \\
2012 & 0.92 & 0.77 & 1.30 & 0.65 & \textcolor{blue}{1.68} \\
2013 & 2.38 & \textcolor{blue}{2.95} & 1.77 & 2.90 & 2.39 \\
2014 & 1.09 & 1.64 & 1.23 & 1.14 & \textcolor{blue}{1.62} \\
2015 & 0.01 & -0.58 & 0.26 & 0.21 & \textcolor{blue}{0.59} \\
2016 & 0.98 & 0.77 & 0.98 & \textcolor{blue}{1.37} & 1.36 \\
2017 & \textcolor{blue}{2.75} & 1.22 & 1.33 & 2.61 & 1.31 \\
2018 & 0.04 & -0.30 & -0.73 & \textcolor{blue}{0.33} & -0.28 \\
   \hline
\end{tabular}
\end{table}

\begin{figure}[ht]
  \centering
  \caption{The line-plot of out of sample annualised return, volatility, Value at Risk and  annualised risk-adjusted return for S\&P 500, CAPM, Fan's model, Factor Model with horseshoe prior, Factor model with ABOS for Markowitz's weight portfolio in Empirical study.}
\begin{tabular}{cc}
\includegraphics[height=6cm,width=6cm]{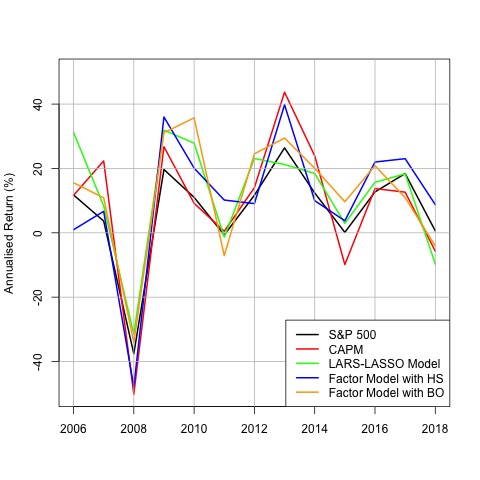}
& \includegraphics[height=6cm,width=6cm]{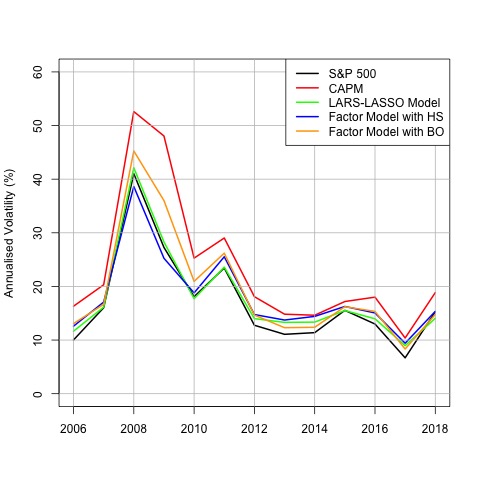}\\
Annualised Return (\%)
& Volatility\\
\hline
\includegraphics[height=6cm,width=6cm]{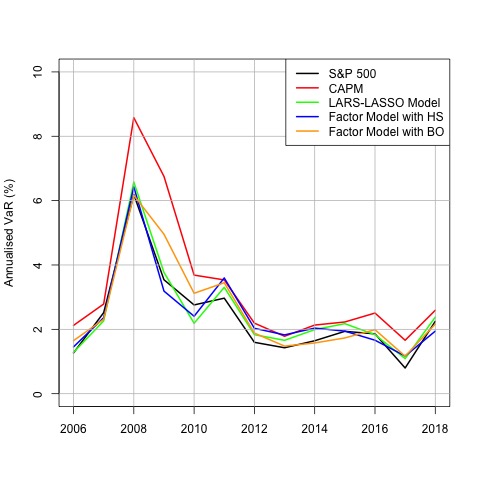}
&\includegraphics[height=6cm,width=6cm]{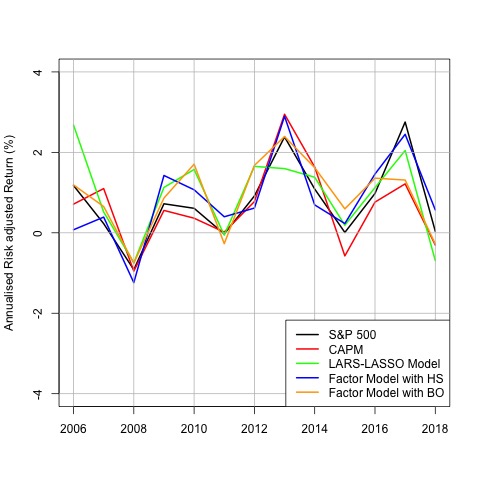}\\
Value at Risk
& Annualised  Risk-adjusted Return\\
\end{tabular}
  \label{Fig_portfolio_anualised_sum_mw_wt}
\end{figure}

\begin{table}[ht]
\centering
\caption{Out-sample Annual Return with Markowitz's Weight.}\label{tbl_return_markowitz_wt}
\begin{tabular}{rrrrrr}
  \hline
 & S\&P 500 & CAPM & LARS-LASSO Model & Factor Model with HS & Factor Model with BO \\
  \hline
2006 & 11.78 & 11.63 & 31.21 & 0.95 & 15.57 \\
  2007 & 3.65 & 22.32 & 8.36 & 6.66 & 10.88 \\
  2008 & -37.58 & -50.16 & -31.41 & -47.88 & -34.04 \\
  2009 & 19.67 & 26.75 & 31.90 & 35.98 & 31.10 \\
  2010 & 11.02 & 9.17 & 27.87 & 20.10 & 35.71 \\
  2011 & -0.68 & 0.49 & -1.36 & 10.18 & -7.10 \\
  2012 & 11.68 & 14.03 & 23.10 & 9.07 & 24.55 \\
  2013 & 26.39 & 43.70 & 21.23 & 39.77 & 29.46 \\
  2014 & 12.39 & 23.91 & 18.44 & 9.99 & 20.07 \\
  2015 & 0.14 & -9.88 & 2.93 & 3.74 & 9.65 \\
  2016 & 12.70 & 13.74 & 15.70 & 21.99 & 20.84 \\
  2017 & 18.42 & 12.63 & 18.32 & 23.04 & 10.94 \\
  2018 & 0.59 & -5.72 & -9.71 & 8.75 & -4.12 \\
   \hline
\end{tabular}
\end{table}

\begin{table}[ht]
\centering
\caption{Annualized volatility of out-sample return of Markowitz's portfolio}\label{tbl_vol_mw}
\begin{tabular}{rrrrrr}
  \hline
 & S\&P 500 & CAPM &  LARS-LASSO Model & Factor Model with HS & Factor Model with BO \\
  \hline
2006 & 10.02 & 16.31 & 11.65 & 12.59 & 13.07 \\
  2007 & 16.02 & 20.30 & 16.14 & 17.08 & 16.66 \\
  2008 & 41.02 & 52.61 & 42.13 & 38.62 & 45.26 \\
  2009 & 27.27 & 48.05 & 28.28 & 25.25 & 36.01 \\
  2010 & 18.10 & 25.28 & 17.71 & 18.80 & 20.94 \\
  2011 & 23.44 & 29.03 & 23.61 & 25.53 & 26.20 \\
  2012 & 12.76 & 18.09 & 13.99 & 14.76 & 14.66 \\
  2013 & 11.07 & 14.82 & 13.30 & 13.73 & 12.31 \\
  2014 & 11.38 & 14.63 & 13.35 & 14.41 & 12.38 \\
  2015 & 15.54 & 17.19 & 15.51 & 16.27 & 16.22 \\
  2016 & 12.99 & 18.00 & 13.96 & 15.05 & 15.34 \\
  2017 & 6.69 & 10.38 & 8.96 & 9.41 & 8.33 \\
  2018 & 15.26 & 18.85 & 14.06 & 15.35 & 14.78 \\
   \hline
\end{tabular}
\end{table}

\begin{table}[ht]
\centering
\caption{Value at Risk (VaR) of out-sample return of Markowitz's portfolio.}\label{tbl_VaR_mw}
\begin{tabular}{rrrrrr}
  \hline
 & S\&P 500 & CAPM &  LARS-LASSO Model & Factor Model with HS & Factor Model with BO \\
  \hline
2006 & 1.27 & 2.12 & 1.29 & 1.45 & 1.65 \\
  2007 & 2.52 & 2.79 & 2.26 & 2.36 & 2.31 \\
  2008 & 6.19 & 8.58 & 6.58 & 6.42 & 6.17 \\
  2009 & 3.54 & 6.75 & 3.77 & 3.19 & 4.96 \\
  2010 & 2.76 & 3.69 & 2.19 & 2.41 & 3.12 \\
  2011 & 2.97 & 3.54 & 3.31 & 3.60 & 3.45 \\
  2012 & 1.60 & 2.19 & 1.83 & 2.03 & 1.88 \\
  2013 & 1.43 & 1.78 & 1.66 & 1.83 & 1.48 \\
  2014 & 1.65 & 2.13 & 1.99 & 2.04 & 1.58 \\
  2015 & 1.94 & 2.23 & 2.18 & 1.94 & 1.73 \\
  2016 & 1.86 & 2.51 & 1.83 & 1.66 & 1.99 \\
  2017 & 0.80 & 1.66 & 1.09 & 1.17 & 1.16 \\
  2018 & 2.25 & 2.59 & 2.38 & 1.94 & 2.16 \\
   \hline
\end{tabular}
\end{table}

\begin{table}[ht]
\centering
\caption{Risk adjusted out-sample return of Markowitz's portfolio. }\label{tbl_sharpe_mw}
\begin{tabular}{rrrrrr}
  \hline
 & S\&P 500 & CAPM &  LARS-LASSO Model & Factor Model with HS & Factor Model with BO \\
  \hline
1 & 1.18 & 0.71 & 2.68 & 0.08 & 1.19 \\
  2 & 0.23 & 1.10 & 0.52 & 0.39 & 0.65 \\
  3 & -0.92 & -0.95 & -0.75 & -1.24 & -0.75 \\
  4 & 0.72 & 0.56 & 1.13 & 1.43 & 0.86 \\
  5 & 0.61 & 0.36 & 1.57 & 1.07 & 1.71 \\
  6 & -0.03 & 0.02 & -0.06 & 0.40 & -0.27 \\
  7 & 0.92 & 0.78 & 1.65 & 0.61 & 1.68 \\
  8 & 2.38 & 2.95 & 1.60 & 2.90 & 2.39 \\
  9 & 1.09 & 1.63 & 1.38 & 0.69 & 1.62 \\
  10 & 0.01 & -0.57 & 0.19 & 0.23 & 0.60 \\
  11 & 0.98 & 0.76 & 1.12 & 1.46 & 1.36 \\
  12 & 2.75 & 1.22 & 2.05 & 2.45 & 1.31 \\
  13 & 0.04 & -0.30 & -0.69 & 0.57 & -0.28 \\
   \hline
\end{tabular}
\end{table}

\end{document}